\documentclass{ifacconf}

\usepackage{graphicx} 
\usepackage{epsfig}   
\usepackage{amsmath}  
\usepackage{amssymb}  
\usepackage{natbib}   
\usepackage{anyfontsize}
\usepackage{calrsfs}
\usepackage{enumitem}
\usepackage{mathdots}
\usepackage{setspace}
\usepackage{color,soul}
\usepackage{algorithm}
\usepackage{algorithmicx}
\usepackage{algpascal}
\DeclareMathAlphabet{\pazocal}{OMS}{zplm}{m}{n}

\newcommand{\norm}[1]{\left\lVert#1\right\rVert}

\makeatletter
\def\endfigure{\end@float}
\def\endtable{\end@float}
\makeatother

\let\ifacconfcaptionwidth\captionwidth
\usepackage[caption=false]{subfig}
\let\captionwidth\ifacconfcaptionwidth

\begin{document}
\begin{frontmatter}

\title{On Low-Rank Hankel Matrix Denoising} 

\thanks[footnoteinfo]{This work is supported by the Swiss National Science Foundation under grant no.: 200021\_178890.}
\thanks[copyright]{This work has been submitted to IFAC for possible publication.}

\author[First]{Mingzhou Yin}
\author[First]{Roy S. Smith}

\address[First]{Automatic Control Laboratory, ETH Z\"{u}rich, Switzerland \\ (e-mail: \{myin,rsmith\}@control.ee.ethz.ch)}

\begin{abstract}                
The low-complexity assumption in linear systems can often be expressed as rank deficiency in data matrices with generalized Hankel structure. This makes it possible to denoise the data by estimating the underlying structured low-rank matrix. However, standard low-rank approximation approaches are not guaranteed to perform well in estimating the noise-free matrix. In this paper, recent results in matrix denoising by singular value shrinkage are reviewed. A novel approach is proposed to solve the low-rank Hankel matrix denoising problem by using an iterative algorithm in structured low-rank approximation modified with data-driven singular value shrinkage. It is shown numerically in both the input-output trajectory denoising and the impulse response denoising problems, that the proposed method performs the best in terms of estimating the noise-free matrix among existing algorithms of low-rank matrix approximation and denoising.
\end{abstract}


\end{frontmatter}

\section{Introduction}

In system identification and signal processing, it is often assumed that the underlying systems or the noise-free signals have certain low-order structures. The objective is then to find an estimate with the particular low-order structure from noisy data, that obtains the closest match to the unknown system or signal. Such low-order structures can often be expressed as rank deficiency in structured data matrices. Thus, the problem of finding a low-order estimate from data can be interpreted as finding structured low-rank matrices from noisy data matrices, known as the structured low-rank approximation (SLRA) problem. See \cite{Markovsky_2008} for an overview.

The problem of estimating an unknown low-rank matrix from noisy measurements has been a long-standing problem, under the name of principal component analysis (\cite{abdi2010principal}) or proper orthogonal decomposition (\cite{Berkooz_1993}). A well-known technique to solve this problem is the truncated singular value decomposition (TSVD), which approximates the data matrices by only keeping the most significant singular values corresponding to the true rank of the noise-free matrix. According to the Eckart-Young-Mirsky (EYM) theorem (\cite{Eckart_1936}), it is the best low-rank approximation to the data in terms of both the Frobenius norm and the spectral norm. When the true rank of the underlying low-rank matrix is unknown, it is still common to rely on the EYM theorem by assuming the rank or turning to the problem of estimating the true rank, e.g., \cite{Bauer_2001} in subspace identification. The simplest method is to look for a sudden decrease in the scree plot (a plot of singular values in decreasing order). Information criteria (\cite{Akaike_1974}) and cross-validation techniques (\cite{STOICA_1986}) are also widely used.

However, an often neglected aspect of the EYM theorem is that it only provides the optimal low-rank approximation to the noisy data matrix, but does not guarantee any optimality of estimating the noise-free matrix (\cite{Nadakuditi_2014}). To estimate the noise-free matrix, one is interested in minimizing the mean squared error (MSE) of the estimate with respect to the noise-free matrix. In this paper, this problem is referred to as the low-rank denoising problem to differentiate from the SLRA problem. 

To solve the low-rank denoising problem, the two-step approach of first estimating the true rank and then applying TSVD can be interpreted as singular value hard thresholding (SVHT), where the singular values are not truncated to obtain a fixed rank but according to a fixed threshold. This method is known to be effective in multiple matrix estimation problems (\cite{Chatterjee_2015}). \cite{Gavish_2014} show that, for unstructured matrices, there exists an optimal choice of the hard threshold asymptotically, which is also effective with finite-dimensional matrices numerically. The hard thresholding function can be generalized to a general shrinkage function on the singular values. The asymptotically optimal shrinkage function for unstructured matrices is developed in \cite{Gavish_2017}. Data-driven and adaptive shrinkage algorithms are also proposed in \cite{Nadakuditi_2014,Josse_2015}.

Another difficulty in exploiting the low-rank prior in estimation is how to incorporate the structural constraints. Regarding the SLRA problem, there is no closed-form solution or convex formulation in general. In existing works, nonlinear optimization algorithms (\cite{Markovsky_2013,park1999low}), iterative structural approximation (\cite{Wang_2019,Ye_Li_1997}), and convex relaxation (\cite{Fazel_2001,Smith_2014}) are applied to obtain locally optimal or suboptimal solutions to the problem. Structure constraints pose additional difficulties in solving the low-rank denoising problem as well. The aforementioned results for the denoising problem rely on the asymptotic distribution of the singular values of the noise matrix, which is not well studied for most structured matrices. In this regard, \cite{Nadakuditi_2014} proposes a data-driven method to estimate the distribution from the singular values of the data matrix.

In this paper, we focus on a generalized Hankel structure, where the underlying low-rank matrix is assumed to be a transformed Hankel matrix. This structure includes, for example, standard Hankel matrices, Toeplitz matrices, and Hankel matrices with known rows/columns. A similar structure is also considered in \cite{Markovsky_2014}. Such structure can be found in various fields, including subspace system identification (\cite{Fazel_2013}), behavioural system modelling (\cite{markovsky2006exact}), reduced-rank signal processing (\cite{Scharf_1991}), and image processing (\cite{Kyong_Hwan_Jin_2015}). In particular, two characteristic examples in linear time-invariant (LTI) systems are investigated. The first one is input-output trajectory denoising, which is used in both time-domain subspace identification (\cite{Moonen_1989}) and behavioral system modelling (\cite{markovsky2006exact}). The second one is impulse response denoising, which is a common step in frequency-domain subspace identification (\cite{McKelvey_1996}) and model order reduction (\cite{Markovsky_2005}).

This paper first reviews existing algorithms in solving the SLRA and the low-rank denoising problem. It is observed that, when applied to the problem of denoising low-rank generalized Hankel matrices, these two categories of algorithms improve the standard TSVD approach from two distinct perspectives, namely enforcing structural constraints and avoiding approximating the noise matrix. Based on this observation, a novel algorithm is then proposed to address the low-rank Hankel matrix denoising problem directly. It combines the data-driven singular value shrinkage approach in unstructured low-rank matrix denoising and the iterative structural approximation method in SLRA. Since rigorous statistical frameworks for low-rank Hankel matrix denoising have not been established, this paper focuses on a numerical analysis perspective to assess the performance in terms of noise reduction by Monte Carlo simulation. It is shown numerically that, when applying to matrices with the generalized Hankel structure, the proposed algorithm achieves the largest noise reduction among all existing SLRA and low-rank denoising algorithms in both examples under different noise levels.

\section{Problem Statement}

For a set of structured matrix $\mathbb{M}^{m\times n}\subseteq\mathbb{R}^{m\times n}$, consider the problem of estimating an unknown matrix $X\in\mathbb{M}^{m\times n}$ from a noisy measurement
\begin{equation}
    W = X+\sigma Z,
\end{equation}
where $\sigma$ is the noise level, $Z\in\mathbb{M}^{m\times n}$ is a stochastic noise matrix with zero mean. It is known that the unknown matrix $X$ has the following low-rank property:
\begin{equation}
    \text{rank}(X\Pi)=r,
\end{equation}
where $\Pi\in\mathbb{R}^{n\times n}$ is a known transformation matrix. Without loss of generality, let $r<m\leq n$, $\beta=m/n\in (0,1]$.

To obtain the optimal estimate of the noise-free matrix $X$, we are interested in minimizing the MSE of the estimate:
\begin{equation}
   \text{MSE}(\hat{X}):=\mathbb{E}\left(\norm{X-\hat{X}}_F^2\right),
    \label{eqn:1}
\end{equation}
where the estimate $\hat{X}$ is a function of the measurement $W$. This problem will be referred to as the denoising problem.

In this paper, we are interested in the case where $\mathbb{M}^{m\times n}$ is the set of $m$-by-$n$ Hankel matrices. By choosing different transformations $\Pi$, this problem formulation covers a class of generalized low-rank Hankel structures including standard Hankel matrices ($\Pi=\mathsf{I}_n$), Toeplitz matrices ($(\Pi)_{i,j}=\mathbf{1}_{\{i+j=n+1\}}$)
and Hankel matrices with noise-free rows ($\Pi$ spans the null space of the noise-free rows).

\subsection{Examples in Linear System Theory}
\label{sec:21}

Two examples of the low-rank matrix denoising problem in LTI systems are investigated. Consider a discrete-time finite-dimensional single-input single-output LTI system:
\begin{equation}
    y_k = G(z)u_k,
\end{equation}
where $u_k$ and $y_k$ are the input and output of the system respectively, and $G(z)$ is an unknown discrete-time transfer function of order $n_x$.

In the first example, a length-$N$ input-output trajectory $(u_k, y_k)_{k=0}^{N-1}$ of the system is measured, where the output measurements are contaminated with additive noise: $\hat{y}_k=y_k+v_k$. Assuming the input trajectory is persistently exciting of order $(m+n_x)$, construct the following mosaic Hankel matrix from the noise-free trajectory:
\begin{equation}
D=
\begin{bmatrix}U\\Y\end{bmatrix}:=\begin{bmatrix}
    u_0&u_1&\cdots&u_{n-1}\\
    \vdots&\vdots&\ddots&\vdots\\
    u_{m-1}&u_m&\cdots&u_{N-1}\\\hline
    y_0&y_1&\cdots&y_{n-1}\\
    \vdots&\vdots&\ddots&\vdots\\
    y_{m-1}&y_m&\cdots&y_{N-1}\\
    \end{bmatrix}\in\mathbb{R}^{2m\times n},
\end{equation}
where $m>n_x$, $n=N-m+1$. Then the signal matrix $D$ has a rank-deficient structure with $\text{rank}(D) = m+n_x$ (\cite{Moonen_1989}).

The output measurements can be denoised based on the low-rank structure of $D$ by constructing $\hat{Y}$ similar to $Y$ but with $(\hat{y}_k)_{k=0}^{N-1}$ instead of $(y_k)_{k=0}^{N-1}$. Thus, the low-rank Hankel matrix denoising problem can be formulated with 
\begin{equation}
    X=Y,\ \,W=\hat{Y},\ \,\Pi=\Pi_U^\perp,\ \,r=n_x
\end{equation}
where $\Pi_U^\perp$ spans the null space of $U$ and can be calculated as $\Pi_U^\perp=\mathsf{I}_n-U^\top(UU^\top)^{-1}U$. The denoised output signal matrix $\hat{X}$ can be used to construct state-space realizations (\cite{Moonen_1989}) or data-driven models (\cite{markovsky2006exact}) of the system. This example will be called input-output trajectory denoising in what follows.

In the second example, the first-$N$ impulse response coefficients $(g_k)_{k=0}^{N-1}$ of the system are measured with additive noise: $\hat{g}_k=g_k+v_k$. Similar to $Y$, construct $m$-by-$n$ Hankel matrices with $(g_k)_{k=0}^{N-1}$ and $(\hat{g}_k)_{k=0}^{N-1}$ and denote them by $H_g$ and $H_{\hat{g}}$ respectively. The matrix $H_g$ is rank-deficient with $\text{rank}(H_g)=n_x$ (\cite{fazel2003log}). This can be seen as a special case of input-output trajectory denoising with an impulse as input. This leads to the denoising problem with
\begin{equation}
    X=H_g,\ \,W=H_{\hat{g}},\ \,\Pi=\mathsf{I}_n,\ \,r=n_x.
\end{equation}
This special case has particular applications in frequency-domain subspace identification (\cite{McKelvey_1996}) and model order reduction (\cite{Markovsky_2005}). This example will be called impulse response denoising in what follows.

\section{Structured Low-Rank Approximation}

Since the MSE depends on the unknown matrix $X$, it is hard to solve the denoising problem directly. Practically, the estimation problem is usually reformulated as finding the best structured rank-$r$ approximation of the measurement $W$:
\begin{equation}
    \begin{matrix}
    \hat{X}_\text{SLRA}=&\underset{\hat{X}\in\mathbb{M}^{m\times n}}{\text{argmin}}&\norm{W-\hat{X}}_F^2\\
    &\text{s.t.}&\text{rank}(\hat{X}\Pi)\leq r.
    \end{matrix}
    \label{eqn:2}
\end{equation}
This problem will be referred to as the approximation problem. The most well-known method to solve this approximation problem is probably TSVD. Let
\begin{equation}
    W\Pi=\sum_{i=1}^m w_i\mathbf{u}_i\mathbf{v}_i^\mathsf{T},
\end{equation}
be the singular value decomposition (SVD) of $W\Pi$, where $w_i$ are the singular values in decreasing order and $\mathbf{u}_i\in\mathbb{R}^m$, $\mathbf{v}_i\in\mathbb{R}^n$ are the left and right singular vectors respectively. Then the TSVD estimate is given by
\begin{equation}
    \hat{X}_\text{TSVD}(W,\Pi;r) = \sum_{i=1}^r w_i\mathbf{u}_i\mathbf{v}_i^\mathsf{T}+W(\mathsf{I}_n-\Pi).
    \label{eqn:eym}
\end{equation}
For the unstructured case, i.e., $\mathbb{M}^{m\times n}=\mathbb{R}^{m\times n}$, $\Pi=\mathsf{I}_n$, the EYM theorem (\cite{Eckart_1936}) shows that $\hat{X}_\text{TSVD}$ is the closed-form solution of (\ref{eqn:2}). This total least squares solution is also the maximum likelihood estimator when $Z$ consists of i.i.d Gaussian entries. 
 
When the matrix is structured, closed-form solutions no longer exist in general, so relaxations or nonlinear optimization techniques are needed to solve the problem. Here, we highlight an algorithm for solving the Hankel low-rank approximation problem by iterating the TSVD step and a Hankel approximation step in an alternating fashion. This method is an extension of the algorithms in \cite{Wang_2019,Ye_Li_1997} to generalized Hankel structure. The algorithm is outlined in Algorithm~\ref{al:1}.

\begin{algorithm}[htb]
	\caption{Iterative algorithm for SLRA with generalized Hankel structure}
	\begin{algorithmic}[1]
	\State \textbf{Input: }$W,\Pi,r,\epsilon$.
	\State $W_1 \leftarrow W$
    \Repeat
        \State $W_2 \leftarrow \hat{X}_\text{TSVD}(W_1,\Pi;r)$
        \State $W_1 \leftarrow \mathcal{H}(W_2)$
    \Until{\norm{W_1-W_2}<\epsilon\norm{W_1}}
	\State \textbf{Output: }$\hat{X}=W_1$.
	\end{algorithmic}
	\label{al:1}
\end{algorithm}

In Algorithm~\ref{al:1}, $\mathcal{H}(\cdot)$ is the orthogonal projector onto the set of Hankel matrices. It can be calculated by setting all the elements along a skew diagonal to be the average value of that skew diagonal.

In addition, two other algorithms proposed in existing literature to solve the SLRA problem are considered on the generalized Hankel structure. The first algorithm, proposed in \cite{Markovsky_2013}, decomposes the optimization problem into a least-norm inner problem and a nonlinear outer problem and solves it by local optimization methods. The second algorithm applies the nuclear norm heuristic of the rank constraint and formulates a regularized convex optimization problem (\cite{Fazel_2001}):
\begin{equation}
    \hat{X}_\text{nuc}=\underset{\hat{X}\in\mathbb{M}^{m\times n}}{\text{argmin}}\ \frac{1}{2}\norm{W-\hat{X}}_F^2+\tau \norm{\hat{X}\Pi}_*,
    \label{eqn:nuc}
\end{equation}
where $\norm{\cdot}_*$ denotes the nuclear norm which is defined as the sum of all singular values. For the unstructured case, it has a closed-form solution of soft-thresholding the singular values:
\begin{equation}
    \hat{X}_\text{nuc}=\sum_{i=1}^m \max(0,w_i-\tau)\mathbf{u}_i\mathbf{v}_i^\mathsf{T}.
\end{equation}

\section{Towards Low-Rank Matrix Denoising with Generalized Hankel Structure}

Despite its wide applications, the SLRA solution $\hat{X}_\text{SLRA}$ to the approximation problem does not always serve as a reasonable solution to the denoising problem. Consider the case when $\sigma\rightarrow\infty$. The optimal denoising solution is a zero matrix almost surely as the low-rank matrix is overwhelmed by noise, whereas $\hat{X}_\text{SLRA}$ approaches infinity which gives only a low-rank approximation of the particular noise realization. This observation illustrates an important aspect of solving the denoising problem: the noise matrix does not only contaminate the left null space of $X\Pi$ and inflate the zero singular values, but also enters the column space of $X\Pi$ and inflates the non-zero singular values.

In detail, let the singular values of $X\Pi$ be $x_i$, $i=1,\dots,m$, where $x_i=0$ for $i>r$. Consider the asymptotic framework where $n\rightarrow\infty$ while keeping both the aspect ratio $\beta$ and the true singular values $x_i$ constant. As proved in Theorem~2.9 of \cite{Benaych_Georges_2012}, the $r$ largest singular values of $W\Pi$ satisfy
\begin{equation}
    \lim_{n\rightarrow\infty}w_i=\begin{cases}D^{-1}_{\mu_Z}(1/x_i^2),&x_i^2>1/D_{\mu_Z}(b^+)\\b,&x_i^2\leq 1/ D_{\mu_Z}(b^+)\end{cases},
    \label{eqn:singpert}
\end{equation}
for $i=1,\dots,r$ almost surely, where $\mu_Z$ is the asymptotic probability measure of the empirical singular value distribution of $Z\Pi$:
\begin{equation}
    \mu_Z = \lim_{n\rightarrow\infty}\frac{1}{m}\sum_{i=1}^m \delta_{z_i},\ \,z_i\text{: singular values of }Z\Pi,
\end{equation}
$b$ is the supremum of the support of $\mu_Z$, and $D_{\mu_Z}(\cdot)$ is the D-transform under $\mu_Z$ (\cite{Benaych_Georges_2012}). An important property of (\ref{eqn:singpert}) is that the noisy singular values are always enlarged, i.e., $w_i>x_i$, $\forall x_i$. Therefore, in addition to setting the smallest $(m-r)$ singular values to zero, the $r$ largest singular values of $W\Pi$ need to be shrunk as well, depending on the singular value distribution of the noise matrix. So for the denoising problem, it makes sense to consider the following singular value shrinkage estimate:
\begin{equation}
\hat{X}_\text{shrink} = \sum_{i=1}^m \eta(w_i)\mathbf{u}_i\mathbf{v}_i^\mathsf{T}+W(\mathsf{I}_n-\Pi),\ \,\eta(w_i)\in[0,w_i]
\label{eqn:shrinkl}
\end{equation}
to counteract the effect of inflated noisy singular values.

In the following subsections, we will start from the unstructured denoising problem, and extend the algorithm to the Hankel matrix denoising problem, where both the noise-free matrix and the noise model are structured.

\subsection{The Unstructured Denoising Problem}
\label{sec:41}

For the unstructured case, it can be assumed that $Z$ has i.i.d. unit Gaussian entries. Then, $\mu_Z$ is known to follow the Marchenko-Pastur distribution (\cite{Mar_enko_1967}). In this case, it has been proven by \cite{Gavish_2014,Gavish_2017} that the following shrinkage law obtains the minimum asymptotic MSE:
\begin{equation}\resizebox{\columnwidth}{!}{$
    \eta(w)=\begin{cases}\frac{n\sigma^2}{w}\sqrt{\left(\frac{w^2}{n\sigma^2}-\beta-1\right)^2-4\beta},&w>(1+\sqrt{\beta})\sqrt{n}\sigma\\0,&w\leq(1+\sqrt{\beta})\sqrt{n}\sigma\end{cases}.$}
    \label{eqn:law1}
\end{equation}
In addition to the general shrinkage function (\ref{eqn:shrinkl}), particular shrinkage functions with piecewise linear forms are often considered. These include hard thresholding and soft thresholding functions, which are defined as
\begin{equation}
    \eta_{H}(w)=w\mathbf{1}_{\{w\geq\tau_H\}},\ \eta_{S}(w)=\max(0,w-\tau_S).
\end{equation}
These functions correspond to TSVD with rank estimation and the nuclear norm regularization for the unstructured case. The optimal thresholds are
\begin{align}
    \tau_H&=\sqrt{2(\beta+1)+\frac{8\beta}{\beta+1+\sqrt{\beta^2+14\beta+1}}}\sigma\sqrt{n},
    \label{eqn:law2}\\
    \tau_S&=(1+\sqrt{\beta})\sigma\sqrt{n},\label{eqn:law4}
\end{align}
respectively. Interestingly, these asymptotically optimal results do not require any knowledge of the true rank $r$. For a comparison of these shrinkage functions, see Figure~2 in \cite{Gavish_2017}.

When the noise level $\sigma$ is unknown, it can be estimated by comparing the last $(m-r)$ singular values of $W\Pi$, which are dominated by noise, to the measure $\mu_Z$. In this work, we apply a robust and consistent estimator proposed in \cite{Gavish_2014}:
\begin{equation}
    \hat{\sigma}=\frac{w_\text{med}}{\sqrt{n\cdot z_\text{med}(\beta)}},
    \label{eqn:sigest}
\end{equation}
where $w_\text{med}$ is the median singular value and $z_\text{med}(\beta)$ is the median of the Marchenko-Pastur distribution.

\subsection{Denoising with Generalized Hankel Noise Model}

When $Z$ is Hankel, the assumption of i.i.d. Gaussian entries in the previous subsection is violated. If the alternative probability measure $\mu_Z$ is known, the optimal shrinkage law (\ref{eqn:law1}) can be extended as
\begin{equation}
    \eta(w;\mu_Z)=\begin{cases}-2\frac{D_{\mu_Z}(w)}{D'_{\mu_Z}(w)},&D_{\mu_Z}(w)<D_{\mu_Z}(b^+)\\0,&D_{\mu_Z}(w)\geq D_{\mu_Z}(b^+)\end{cases},
\end{equation}
according to Theorem~2.1 in \cite{Nadakuditi_2014}. Unfortunately, to the best of our knowledge, the empirical singular value distribution of random Hankel matrices has only been analyzed numerically (e.g. \cite{Ghodsi_2015,Smith_2014}) but lacks an analytical formulation.

So, instead of aiming to derive the optimal shrinkage law analytically, the data-driven singular value shrinkage algorithm, \textit{OptShrink}, can be applied (Algorithm~1 in \cite{Nadakuditi_2014}). This algorithm obtains a consistent estimate of the measure $\mu_Z$ of the noise singular value distribution from the last $(m-r)$ singular values of $W\Pi$. It can be considered as an extension of the noise level estimator (\ref{eqn:sigest}). In the unstructured case, the distribution has been parametrized by noise level $\hat{\sigma}$ with the Marchenko-Pastur distribution. Here, a nonparametric estimation of the empirical singular value distribution is obtained. So the data-driven shrinkage law is given by
\begin{equation}
    \eta_\text{DD}(w_i)=\begin{cases}\eta(w_i;\hat{\mu}_Z(w_{r+1},\dots,w_{m})),&i=1,\dots,r\\0,&i=r+1,\dots,m\end{cases}.
    \label{eqn:law3}
\end{equation}
Note that this algorithm requires knowledge of the true rank $r$ to distinguish the singular values that are only resulted from noise.

\subsection{Enforcing the Generalized Hankel Structure}

In addition to the problem with the generalized Hankel noise model, the previous algorithms to solve the denoising problem also do not guarantee the Hankel structure of the unknown matrix $X$. To enforce the Hankel structure in the denoised estimate, we modify Algorithm~\ref{al:1} by replacing the TSVD solution with the data-driven singular value shrinkage law (\ref{eqn:law3}) as follows:
\begin{algorithm}[htb]
	\caption{Iterative algorithm for low-rank denoising with generalized Hankel structure}
	\begin{algorithmic}[1]
	\State \textbf{Input: }$W,\Pi,r,\epsilon$.
	\State $W_1 \leftarrow W$
    \Repeat
        \State 
        $W_2 \leftarrow \sum_{i=1}^r \eta(w_i;\hat{\mu}_Z(w_{r+1},\dots,w_{m}))\mathbf{u}_i\mathbf{v}_i^\mathsf{T}+W_1(\mathsf{I}_n-\Pi),$
        where $(w_i)_{i=1}^m$ are the singular values of $W_1\Pi$.
        \vspace{0.2em}
        \State $W_1 \leftarrow \mathcal{H}(W_2)$
    \Until{\norm{W_1-W_2}<\epsilon\norm{W_1}}
	\State \textbf{Output: }$\hat{X}=W_1$.
	\end{algorithmic}
	\label{al:2}
\end{algorithm}

\section{Numerical Results}

In this section, we compare numerically the performance of the algorithms discussed in the previous sections on the two examples of low-rank Hankel matrix denoising discussed in Section~\ref{sec:21}, namely the input-output trajectory denoising and the impulse response denoising problems. The algorithms are summarized as follows.
\begin{enumerate}
\item Truncated singular value decomposition (\textit{TSVD}): Equation~(\ref{eqn:eym})
\item SLRA by iteration (\textit{Iter}): Algorithm~\ref{al:1} with $\epsilon=10^{-5}$
\item SLRA by local optimization (\textit{SLRA}): SLRA package (\cite{Markovsky_2014})
\item Nuclear norm regularization (\textit{Nuc}): convex optimization problem~(\ref{eqn:nuc}), $\tau$ is selected as the optimal soft threshold (\ref{eqn:law4})
\item Optimal shrinkage law (\textit{Shrink}): Equation~(\ref{eqn:shrinkl}) with shrinkage law (\ref{eqn:law1})
\item Optimal hard thresholding (\textit{Hard}): Equation~(\ref{eqn:shrinkl}) with hard threshold (\ref{eqn:law2})
\item Data-driven shrinkage law (\textit{DD}): Equation~(\ref{eqn:shrinkl}) with data-driven shrinkage law (\ref{eqn:law3})
\item Iterative low-rank Hankel matrix denoising (\textit{LRHD}): Algorithm~\ref{al:2} with $\epsilon=10^{-5}$
\end{enumerate}
In these methods, (2)--(4) are SLRA methods, (5)--(7) are unstructured matrix denoising methods, and (8) is the proposed method. When needed, the true rank $r$ is assumed known.

In both examples, random fourth-order systems generated by the \texttt{drss} function in \textsc{Matlab} are considered ($n_x=r=4$). The number of rows $m$ is selected as 8. The additive noise $v(k)$ is considered as i.i.d. Gaussian noise with $\mathcal{N}(0, \sigma_v^2)$. The noise level $\sigma_v$ is assumed unknown for the algorithms. In the input-output trajectory denoising example, the trajectory length is selected as $N=96$. Two different noise levels of $\sigma_v^2=0.1$ and $0.01$ are considered. In the impulse response denoising example, a shorter length of $N=40$ is selected since the impulse response decays exponentially for stable systems. Two different noise levels of $\sigma_v^2=0.01$ and $0.001$ are considered.

The performance is assessed by the following noise reduction measure:
\begin{equation}
    F=100\cdot\left(1-\dfrac{\norm{X-\hat{X}}_F}{\norm{X-W}_F}\right),
\end{equation}
where $F=0$ means no noise reduction and $F=100$ means the noise-free matrix is fully recovered. For each test case, 100 Monte Carlo simulations are conducted.

\begin{figure}[htbp]
\centering
\includegraphics[width=\columnwidth]{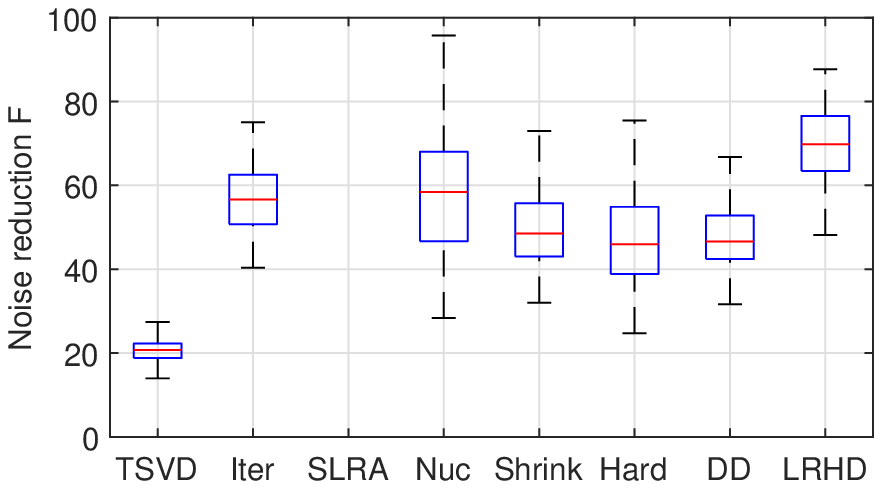}
\footnotesize (a) $\sigma_v^2=0.1$
\includegraphics[width=\columnwidth]{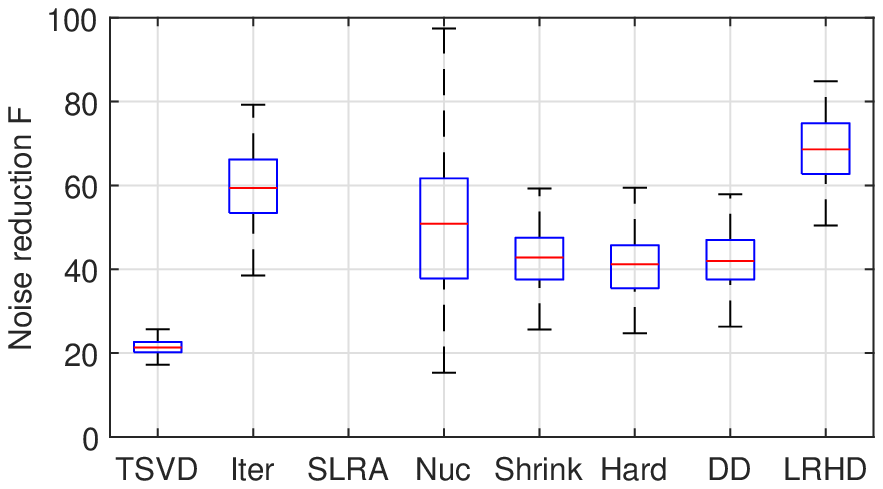}
\footnotesize (b) $\sigma_v^2=0.01$
\caption{Noise reduction performance for the input-output trajectory denoising problem.}
\label{fig:1}
\end{figure}

\begin{figure}[htbp]
\centering
\includegraphics[width=\columnwidth]{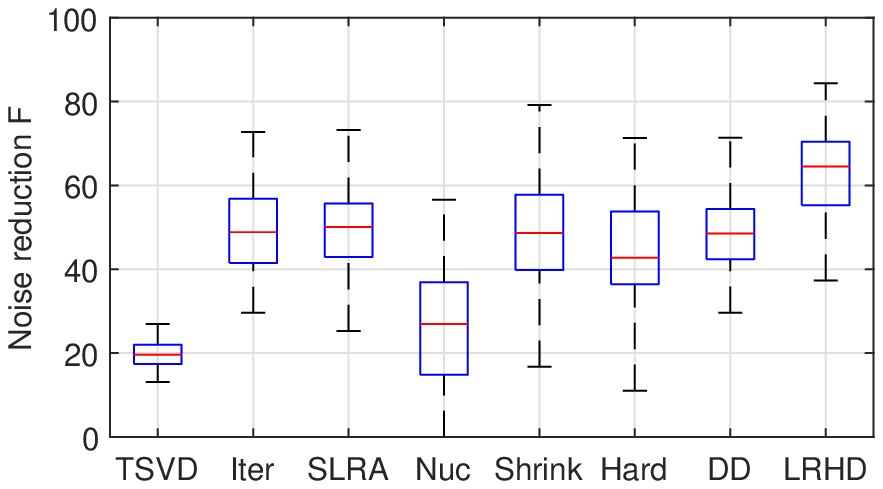}
\footnotesize (a) $\sigma_v^2=0.01$
\includegraphics[width=\columnwidth]{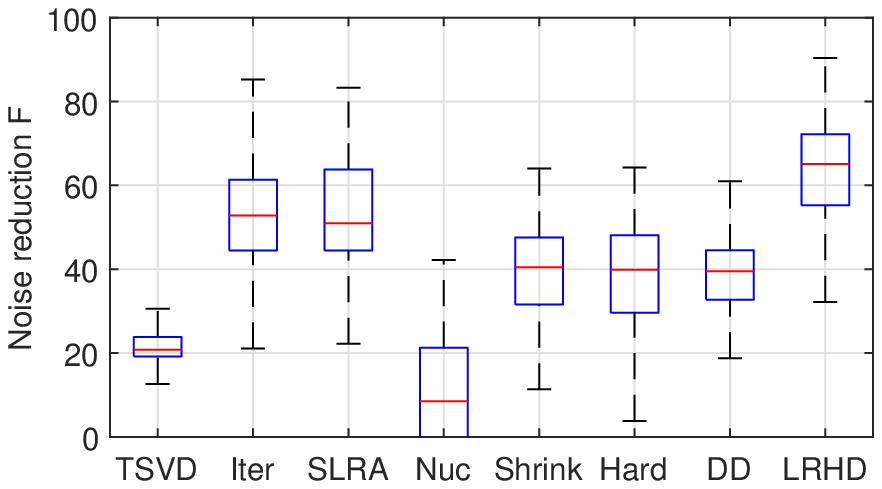}
\footnotesize (b) $\sigma_v^2=0.001$
\caption{Noise reduction performance for the impulse response denoising problem.}
\label{fig:2}
\end{figure}

The boxplots of the noise reduction measure $F$ are plotted in Figures~\ref{fig:1} and \ref{fig:2} for both examples respectively. It can be seen that the proposed iterative low-rank Hankel matrix denoising algorithm achieves the largest noise reduction in both examples at different noise levels. This result proves the benefit of combining the asymptotically optimal singular value shrinkage law with structural constraints. For the most part, the other SLRA and matrix denoising algorithms also perform much better than the TSVD approach, except that \textit{SLRA} fails to obtain a reasonable solution for the input-output trajectory denoising case and \textit{Nuc} does not perform well in the impulse response denoising problem. This demonstrates that despite being the optimal low-rank approximation for unstructured matrices, the performance of TSVD is not satisfying in terms of estimating the structured low-rank matrix.

\section{Conclusion}

In this paper, we have proposed a new approach to the low-rank Hankel matrix denoising problem. Instead of aiming to find a low-rank approximation of the noisy matrix, the approach applies the singular value shrinkage law that is asymptotically optimal in terms of estimating the noise-free matrix. Together with an iterative algorithm to enforce the generalized Hankel structure, this algorithm achieves the best noise reduction performance numerically compared to other low-rank approximation or denoising algorithms.

\bibliography{refs}             

\end{document}